\documentclass[lettersize,journal]{IEEEtran}
\IEEEoverridecommandlockouts
\usepackage{cite}
\usepackage{epstopdf}
\usepackage{graphicx}
\usepackage[normalem]{ulem}
\usepackage{subcaption} 
\usepackage{eucal}
\usepackage{amsmath,amssymb,amsfonts}
\usepackage{algorithmic}
\usepackage{graphicx}
\usepackage{textcomp}
\usepackage{ amssymb }
\usepackage{mathrsfs}
\usepackage{algorithm}
\usepackage{xcolor}
\def\BibTeX{{\rm B\kern-.05em{\sc i\kern-.025em b}\kern-.08em
    T\kern-.1667em\lower.7ex\hbox{E}\kern-.125emX}}
\begin{document}

\title{\huge Joint Power Allocation and Radiation Optimization in NOMA-Assisted Pinching Antenna Systems}
\author{Nikoloz Vashakidze, Chadi Assi, Mohamed Elhattab, Ali Ghrayeb and Maurice J. Khabbaz \vspace{-0.5cm}}

\maketitle

\begin{abstract} 

This paper explores a joint optimization of transmit power allocation and radiation coefficients in a downlink Pinching Antenna SyStem (PASS) employing Non-Orthogonal Multiple Access (NOMA). By leveraging the PASS-enabled flexible channel adjustment and NOMA's power allocation adaptability, a sum rate maximization problem is formulated with the objective of simultaneously optimizing base station (BS)'s transmit power coefficients and pinching antenna (PA)'s radiation powers. Due to its non-convexity and complexity, the formulated optimization problem is challenging to solve directly. Hence, we decompose the main problem into two sub-problems, namely transmit power allocation sub-problem and PA radiation power allocation sub-problem. In the first sub-problem, closed-form solutions are derived for the BS's power allocation among NOMA users. Meanwhile, in the second sub-problem, we optimize the PA's radiation power utilizing successive convex approximation (SCA). These two sub-problems are solved alternatively using Alternating Optimization (AO) until convergence. It should be noted that decoding order plays a significant role in NOMA-assisted PASS. Hence, two variations of decoding order are considered, namely: \textit{i}) a high-complexity exhaustive search approach, and, \textit{ii}) a low-complexity alternative that  utilizes pre-determined channel information. Numerical results show that our proposed approach substantially improves the system's sum-rate compared to widely adopted equal power allocation PASS schemes.
\end{abstract}

\section{Introduction}
Wireless communication technologies are undergoing a period of rapid transformation driven by the vision and requirements of future 6G cellular networks, \cite{10054381}. Next-generation systems aim at delivering unprecedented levels of automation, intelligence, connectivity, and data rate enhancements intended to support emerging and disruptive services outlined by 3rd Generation Partnership Project (3GPP), International Mobile Telecommunications (IMT), and European Telecommunications Standards Institute (ETSI) standards, \cite{10904090}. The evolution to 6G will enable a diverse spectrum of advanced applications that leverage high-capacity networking, ultra-low latency, and seamless integration of devices, empowering new capabilities for intelligent environments and digital services, \cite{10054381}.

Modern 5G and emerging 6G wireless networks are underpinned by a suite of advanced technologies that transform connectivity and signal processing. Key enablers (\textit{e.g.}, Multiple-Input Multiple-Output (MIMO) and massive MIMO, Non-Orthogonal Multiple Access (NOMA), Rate-Splitting Multiple Access (RSMA), as well as Reconfigurable Intelligent Surfaces (RIS) and Movable Antenna (MA) systems) collectively drive unprecedented gains in capacity, energy efficiency, and user density \cite{liu2025pinchingTutorial}. Complementing these, Pinching Antenna System (PASS) has recently emerged as a novel technology with advantages over their counterparts, \cite{11169486}. By deploying dielectric waveguides extending over tens of meters and integrating selectively positioned pinches, these systems allow for dynamic signal radiation and precise control over propagation \cite{liu2025pinchingTutorial}. Through elementary Pinching Antennas' (PAs) adjustments, PASS uniquely enables the creation and targeted manipulation of Line-of-Sight (LoS) channels; thus, further enhancing the flexibility and performance potential of next-generation wireless architectures \cite{liu2025pinchingTutorial}.

To serve downlink users within PASS, several flexible configurations are possible. One approach is to assign each user a dedicated waveguide equipped with PAs that are selectively activated to serve that specific user \cite{liu2025pinchingTutorial}. Alternatively, a single waveguide can accommodate signals for multiple users via Time Division Multiple Access (TDMA), where the antenna configuration is dynamically adjusted during each time slot to direct transmissions towards the selected user \cite{liu2025pinchingTutorial}. This work focuses on a third strategy in which the Base Station (BS) employs NOMA to superimpose users’ signals in the power domain, enabling all users to be served simultaneously through the same waveguide and shared set of PAs \cite{11215676}. 
At the receiver, each user employs Successive Interference Cancellation (SIC): starting from the weakest user (as determined by effective channel quality), they iteratively decode and subtract signals intended for users with lower channel gains before retrieving their own signal. This SIC-based decoding order, a crucial aspect of the system, typically follows the effective channel strength, as established in recent literature. 
This work focuses on the NOMA-enabled approach, leveraging the unique PASS re-configurability for enhanced multi-user spectral efficiency.
  
Different power radiation modes for PAs have been discussed (\textit{e.g.}, \cite{liu2025pinchingTutorial}), two main examples of which are the General Power Radiation (GPR) model and its specific, simplified case, the Equal Power Radiation (EPR) model. In GPR, by adjusting the PA distances from the waveguide, it is possible to control what fraction of the total power is radiated through each PA. Alternatively, in EPR, the PA distances from the waveguide need to be pre-configured so that each subsequent PA is closer to the waveguide than the last one by a specific distance; hence, allowing the power through pinches to be radiated equally \cite{liu2025pinchingTutorial}. In a EPR system, where PAs are activated/deactivated based on the user distribution and system characteristics (\textit{e.g.}, \cite{AntennaActivation}), for each unique PAs' configuration, the system will require dynamic control over the power radiation coefficient to ensure that the power dissipated from each PA remains equally allocated, even as the configuration of PAs changes. This paper explores the problem of optimally configuring the power radiated from the PAs, as well as the NOMA power allocation for users' signals transmitted through the same waveguide.

 
Substantial work has been done specifically in the scope of PA-assisted NOMA systems because of the natural affinity between the two systems.
For example, \cite{11215676} considers a sum rate maximization problem for two users, by adjusting PA positions along the waveguide and the NOMA power allocation of the user. Similarly, \cite{AntennaActivation} considers static PAs' locations, but manages to maximize sum rate by choosing the optimal PA activation configuration. Additional work regarding NOMA user power allocation and optimal SIC ordering is discussed in \cite{OptimalSIC}. In turn, \cite{liu2025pinchingTutorial} discusses different power models, their advantages and challenges.  
 
This paper focuses on maximizing the sum rate in NOMA-assisted PASS by adopting GPR model for NOMA communication and jointly optimizing the BS's transmit power allocation coefficients for NOMA users and PA radiation power coefficients. By comparing the results with the similar system employing an equal power model, paper demonstrates that the proposed system offers 10\% - 15\% increased sum rate depending on the number of users and PAs   . 
This problem is highly coupled and non-convex, yet, resolvable using Alternating Optimization (AO). Specifically, by first fixing the radiation power, it is possible to derive a closed-form expression for the BS transmit power coefficients for each user. Based on the obtained BS transmit power coefficients, PA radiation powers are optimized by using successive convex approximation (SCA). These two sub-problems are solved alternatively until convergence. Two decoding order strategies are considered, namely: \textit{i}) a high-complexity exhaustive search method, and, \textit{ii}) a low-complexity approach that leverages pre-determined channel information. Extensive simulations highlight the proposed method's significant system sum-rate enhancements compared to commonly used equal power allocation PASS schemes.

The remainder of this paper is organized as follows. The system model is presented in Section II, while Section III lays out the optimization problem's formulation, the resolution of which is addressed in Section IV. Numerical analyses are presented in Section V, and Section VI concludes the paper.
\section{System Model}
This paper focuses on a PASS-assisted downlink NOMA system. A single waveguide with multiple PAs is used to transmit to $K$ single-antenna users uniformly distributed in a 2D rectangular region of side lengths $D_1$ and $D_2$. $h$ is the BS' height. The location of users are given by $\psi^U_k= (x_k,y_k,0), \text{for } k \in \mathcal{K}$, where $\mathcal{K} = [1,2, \dots ,K]$ is the set of all users. To communicate with the users, the BS uses a waveguide equipped with $N_t$ transmit PAs equally spaced along the waveguide's length. Herein, PAs are fixed, and the position of the $n^\text{th}$ pinch is given by $\psi^{PA}_n = (x^P_n, 0, h),  \text{ for } n \in \mathcal{N}$, where $x^P_n$ is the $x$-coordinate of the $n^\text{th}$ pinch and $\mathcal{N} = [1,2,\dots N_t]$ is the set of all PAs. 
\subsection{Pinching Antenna Channel Model}
The signal traveling from the BS located at $\psi^{BS} = (D_1/2 , 0 , h)$ through the waveguide will experience attenuation; the channel from the BS to the $n^{th}$ PA is \cite{AntennaActivation}:
\begin{equation}
    h^{P}_n = 10^{-\frac{k\left(\psi^{BS} -\psi^{PA}_n\right)}{10}}e^{-2\pi j\frac{{\left(\psi^{BS} -\psi^{PA}_n\right)}}{\lambda_g}},
\end{equation}
where $\kappa$ is the waveguide attenuation factor, and $\lambda_g = \frac{\lambda}{n_{\text{eff}}}$ is the wavelength along the wavegide, with $n_{\text{eff}}$ representing the effective refracting index. Using the spherical wave channel model, the channel between user $k$ and the BS, denoted as $\mathbf{h}_k \in \mathbb{C}^{1 \times N_t}$, can be expressed as:
\begin{equation}
    \mathbf{h}_k = \left[h^P_1 \frac{\lambda}{4\pi|\psi^U_k-\psi^{PA}_1|}, \cdots, h^P_{N_t} \frac{\lambda}{4\pi|\psi^U_k-\psi^{PA}_{N_t}|}\right].
\end{equation}

\begin{figure}[!t]
  \centering
  \includegraphics[width=0.8\linewidth]{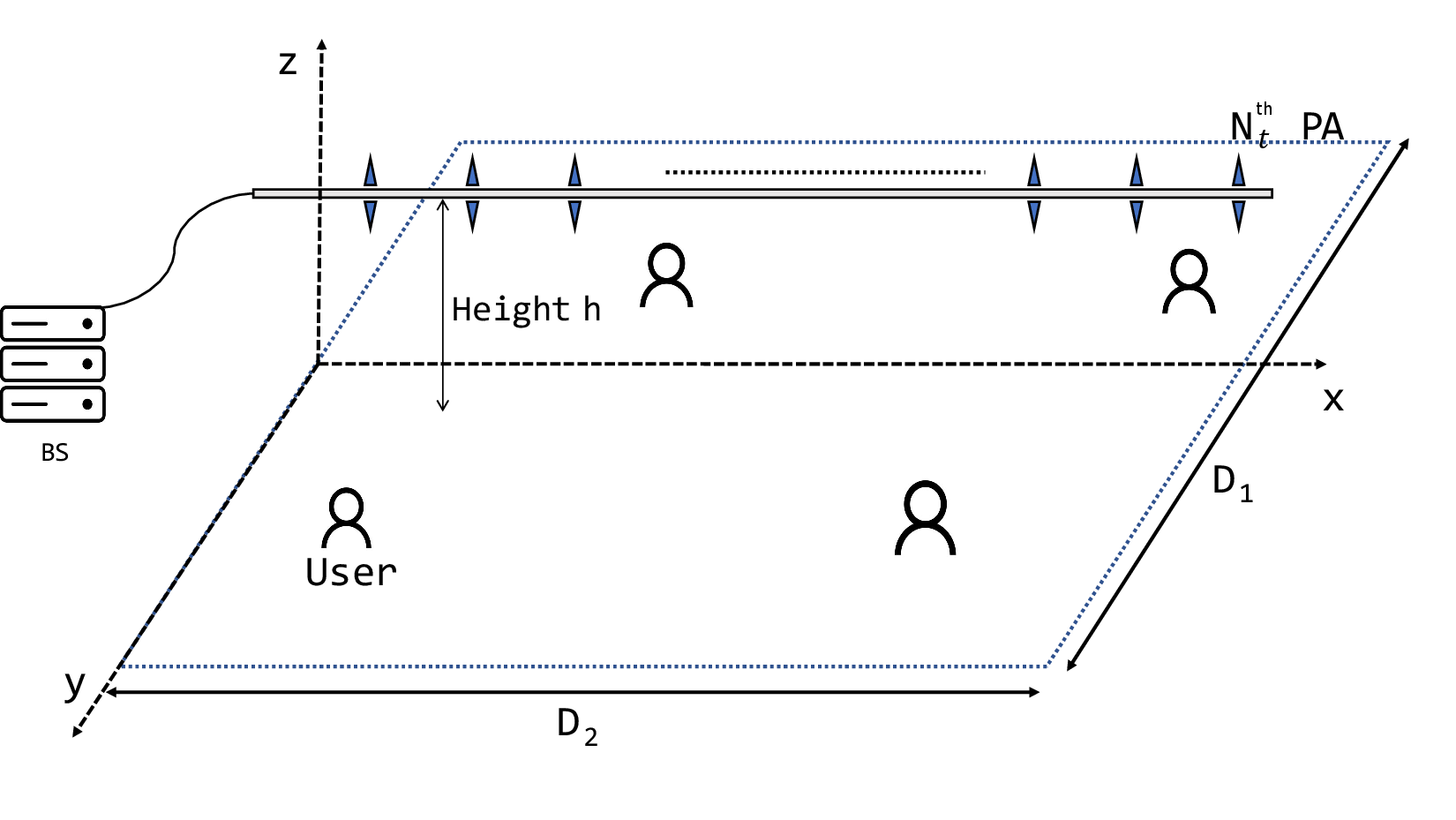}
  \caption{Considered pinching antenna system design}
  \label{fig:system_diagram}
  \vspace{-0.5cm}
\end{figure}

\subsection{Signal Model}
The superimposed signal (through NOMA) sent from the BS is given as $s = \sum_{k =1}^{K}\sqrt{\alpha_k}s_k$. Here, $\alpha_k, \forall k \in \mathcal{K}$ is the fraction of power allocated by the BS to user $k$'s signal (\textit{i.e.}, $s_k$) and its expected value is given by $E[s_k] = 1, \forall k \in \mathcal{K}$. The signal received by user $k$ can be expressed as:
\begin{equation}
    y_k = \mathbf{h}_k\mathbf{p}^{T}s +w_k,
\end{equation}
where $\omega_k \sim C\mathcal{N}(0,\sigma^2) $ is the Additive White Gaussian Noise (AWGN) at user $k$ with zero mean and variance $\sigma^2$. $\mathbf{p} \in \mathbb{R}_{\ge 0}^{1 \times N_t}
 $ antenna radiation power that can also be optimized to maximize the system sum rate. Each element in $\mathbf{p}$ represents the power dissipated from the $n^\text{th}$ PA and $\sum_{n = 1}^{N_t}p_n = P_t$, where $P_t$ is the BS' total emitted power.
\subsection{Power Model}
The system considered in this paper adopts the general and equal power models, which were discussed in \cite{liu2025pinchingTutorial}. The difference between the two models is the level of control the system has over the radiated power from each PA. Adjusting the radiation power at the transmitting PAs offers PASS a new degree of freedom that the optimizer can exploit to achieve a higher rate. To analyze the feasibility of both power models, the PASS' physical properties are examined.
\subsubsection*{\textbf{1) General Power}}
According to the general power radiation model \cite{liu2025pinchingTutorial}, the power radiated from $m^\text{th}$ PA is: 
\begin{equation}
     P_m = \delta^2_m \prod_{i=1}^{m-1}\left(1-\delta_i^2\right),
\end{equation}
where $\delta_m = \sin\left(\kappa_mL_m\right), \forall m .$
%
Here, $\kappa_m = \Omega_0e^{-\sqrt{\gamma_0^2-\frac{4\pi^2}{\lambda^2}n^2_{clad}}S_m}$ and $L_m$ represent the coupling coefficient and coupling length, respectively. Clearly, $\kappa_m$ is a function of several variables, including the spacing $S_m$ between the waveguide and the $m^\text{th}$ PA. By adjusting the PA spacing, $S_m$ or $L_m$, it becomes possible to directly control the fraction of the total power dissipated from any PA. According to \cite{liu2025pinchingTutorial}, the target power radiation ratio, $\beta^{target}_n$, is determined by adjusting the spacing such that: 
\begin{multline}
    \sin\left(\Omega_0e^{-\sqrt{\gamma_0^2-\frac{4\pi^2}{\lambda^2}n^2_{clad}}S_n}L\right) = \\
    \frac{\beta^{target}_n}{\prod_{i=1}^{n-1}\left(1-\sin^2\left(\Omega_0e^{-\sqrt{\gamma_0^2-\frac{4\pi^2}{\lambda^2}n^2_{clad}}S_i}L\right)\right)},
\end{multline}
\subsubsection*{\textbf{2) Equal power}}
One may choose to forgo radiation power control and adopt a configuration in which the power dissipated by each PA is identical. In this case, the power radiated from the $m^\text{th}$ PA is  given by \cite{liu2025pinchingTutorial}:
\begin{equation}
    P_m = P_1 = P_2 = \dots =P_{N_t} \triangleq P_{eq},
\end{equation}
with
\begin{equation}\label{EPM_delta}
    \delta_m = \sqrt{\frac{P_{eq}}{1-(m-1)P_{eq}}}, \forall m \in N_t.
\end{equation}
From \eqref{EPM_delta}, it is observed that, to maintain equal power, the permutation of the PAs must remain unchanged. 
\subsection {SIC Decoding Order and Rate analysis}
In this work, the NOMA decoding order between all users is given by a decoding order vector $\Omega$. Given $\Omega$, user $j$ is scheduled to decode the message of user $k$ iff $\Omega(j) > \Omega(k), \forall k,j \in \mathcal{K}, k\neq j$.
Given the specific case of SIC ordering where $\Omega(k) = k, \forall k\in \mathcal{K}$, user $m$ will first decode the message intended for user $s_1$, treating all other signals as interference. After successfully decoding the first user, the decoded signal gets subtracted from the received signal and user $m$ moves to decode the next weakest message, repeating the process until it reaches $s_m$. Following the above SIC ordering, the rate at which user $j \in \mathcal{K}$ decodes the signal intended for user $m \in \mathcal{K}$ is given by:
\begin{equation}
R_{m \rightarrow{}j} = 
\log_2\!\left(1 + \frac{\left|{\mathbf{h}}_j\mathbf{p}^T \right|^2\alpha_m}
{\sum\limits_{\Omega(i) >\Omega(m)} \left|{\mathbf{h}}_j \mathbf{p}^T  \right|^2\alpha_{i} + \sigma^2}\right),
\end{equation}
With successful SIC, the rate expression at user $j$, then, is:
\begin{equation}
    R_{j} = 
\log_2\!\left(1 + \frac{\left|{\mathbf{h}}_j\mathbf{p}^T \right|^2\alpha_j}
{\sum\limits_{\Omega(i) >\Omega(j)} \left|{\mathbf{h}}_j \mathbf{p}^T  \right|^2\alpha_{i} + \sigma^2}\right).
\end{equation}
To guarantee successful SIC, the rate at user $j$ scheduled to decode user $m$'s signal must be higher than the rate at which user $m$ decodes its own signal \cite{beamforming}), which can be written as follows,%
\begin{equation}\label{R_mj}
R_{m \rightarrow j} \geq R_{m}
\end{equation}
Constraint \eqref{R_mj} guarantees that the BS encodes $s_m$ at a rate lower than what user $j$ can achieve, and therefore decode. Here, the effective channel is a function of the radiation power. To solve this challenge, the algorithm adopts a dynamic SIC ordering. Specifically, after determining the optimal $\mathbf{p}$, it also updates the SIC ordering based on the new effective channel.
\section{Problem Formulation}
The goal is to maximize the sum rate of users by optimizing the joint user NOMA power allocation $\mathbf{\alpha}$ and PA radiation power $\mathbf{p}$ vectors, subject to the power constraint and SIC decoding rate. The optimization problem is formulated as:
\allowdisplaybreaks
\begingroup
\begin{subequations}\label{eq:P1}
\begin{align}
\mathcal{P}1:\quad
&\max_{\mathbf{p}, \boldsymbol{\alpha}} \;\; \sum_{k=1}^K R_{k} \label{eq:P1a}\\
\text{s.t.}\quad
&R_{k\to j} \ge R_{k},\; \Omega(j)>\Omega(k),  \label{eq:P1b}\\
&R_k \ge R_k^{min}, \forall k\in K,\label{eq:P1c} \\
&\sum_{k=1}^{N_t} \|P_k\| \le P_T, \label{eq:P1d}\\
&\sum_{k=1}^{K} \|\alpha_k\| \le 1.\label{eq:P1e}
\end{align}
\end{subequations}
\endgroup
Here, $P_{T}$ denotes the BS's total emitted power. \eqref{eq:P1b} secures a decoding rate that is sufficient for successful SIC. Constraint \eqref{eq:P1c} guarantees the required Quality of Service (QoS). The total radiation power from the PAs is limited by the power budget at the BS \eqref{eq:P1d} and the user NOMA power allocation is bounded by \eqref{eq:P1e}. It can be seen that $\mathcal{P}1$ is non-convex due to the objective function and constraint \eqref{eq:P1b}. To resolve this, $\mathcal{P}1$ is decomposed into two subproblems, one where user NOMA power is optimized for a given radiation power, and, another where radiation power is optimized for a given user NOMA power; hence, allowing AO to get a suboptimal solution. 
\vspace{-0.3cm}
\section{The Proposed Solution}
Hereafter, the above-mentioned two subproblems are resolved, and an overarching algorithm is provided.
\vspace{-0.3cm}
\subsection{User Power Allocation}
To optimize user NOMA power coefficients, we define the following subproblem:
\begin{subequations}\label{eq:P2.1}
\begin{align}
\text{(P2.1):}\quad
&\max_{\boldsymbol{\mathbf{\alpha}}} \;\; \sum_{k=1}^K R_{k}, \label{eq:P21a}\\
\text{s.t.}\quad
&R_{k\to j} \ge R_{k},\; \Omega(j)>\Omega(k), \forall j,k \in \mathcal{K}, \label{eq:P21b}\\
&R_k \ge R_k^{\mathrm{min}}, \forall k\in \mathcal{K},\label{eq:P21c} \\
&\sum_{k=1}^{K} \|\alpha_k\| \le 1. \label{eq:P21d}
\end{align}
\end{subequations}
This problem can be further simplified by examining constraint \eqref{eq:P21b} for a given $\Omega$ and \textbf{p}:
{\footnotesize
\begin{align}
&\log\left( 1 + \frac{|\mathbf{h}_j \mathbf{p}^T|^2 \alpha_k}{\displaystyle \sum_{i=k+1}^{K} |\mathbf{h}_j \mathbf{p}^T|^2 \alpha_i + \sigma^2} \right)
> \notag \\ & \hspace{3cm}
\log\left( 1 + \frac{|\mathbf{h}_k \mathbf{p}^T|^2 \alpha_k}{\displaystyle \sum_{i=k+1}^{K} |\mathbf{h}_k \mathbf{p}^T|^2 \alpha_i + \sigma^2} \right)
\end{align}}
Since the logarithm function is monotonically increasing, the above equation is equivalent to:
\begin{equation}
\frac{|\mathbf{h}_j \mathbf{p}^T|^2 \alpha_k}{\displaystyle \sum_{i=k+1}^{K} |\mathbf{h}_j \mathbf{p}^T|^2 \alpha_i + \sigma^2} 
>
\frac{|\mathbf{h}_k \mathbf{p}^T|^2 \alpha_k}{\displaystyle \sum_{i=k+1}^{K} |\mathbf{h}_k \mathbf{p}^T|^2 \alpha_i + \sigma^2},
\end{equation}
%
%
which leads to: $|\mathbf{h}_j \mathbf{p}^T|^2 > |\mathbf{h}_k \mathbf{p}^T|^2.$

Given the user ordering $\Omega$ such that, $\Omega(j) > \Omega(k)$ iff $|\mathbf{h}_j\mathbf{p}|^2 > |\mathbf{h}_k\mathbf{p}|^2$, \eqref{eq:P21b} is always satisfied. The closed form expression for the optimal $\alpha$ is adapted from \cite{OptimalSIC} as follows: 


%
\begin{align}
\nonumber \alpha_{i}^{\star} &= 
\left[
\beta_{i} 
\left(\prod_{j=1}^{i-1} (1 - \beta_{j})  
+ \frac{1}{\bar{h}_{i}} 
- 
\sum_{j=1}^{i-1}
\frac{
\prod_{k=j+1}^{i-1} (1 - \beta_{k})
\beta_{j}
}{
\bar{h}_{j}
}
\right)
\right]^{+}\\ 
&\qquad\qquad\qquad\qquad\qquad\qquad\quad\forall i = 1, \ldots, K-1
\label{eq:7}
\end{align}
and
\begin{multline}
\alpha_{K}^{\star} =
\left[1-
\sum_{i=1}^{K-1}
\beta_{i}
\left( \prod_{j=1}^{i-1} (1 - \beta_{j}) 
+ \frac{1}{\bar{h}_{i}} \right. \right. \\ \left. \left.
- \sum_{j=1}^{i-1}
\frac{
\prod_{k=j+1}^{i-1} (1 - \beta_{k})
\beta_{j}
}{
\bar{h}_{j}
}
\right)
\right]^{+},
\label{eq:8}
\end{multline}
\text{where} $\beta_{i} = \left(2^{R^{\min}_i} - 1\right)/2^{R^{\min}_i}, \quad \forall i = 1, \ldots, K-1,$ $\bar{h}_n=\frac{h_n}{\sigma^2}$
\text{and} $[\,\cdot\,]^{+} = \max\{\cdot, 0\}.$ \\
\vspace{-0.4cm}
\subsection{Radiation Power Allocation}
After obtaining the user power allocation, the second subproblem (P2.2) is formulated below to optimize the PA radiation power for a given user NOMA power and SIC ordering, which can be formulated as follows,
\begin{subequations}\label{eq:P2.2}
\begin{align}
\text{(P2.2):}\quad
&\max_{\boldsymbol{p}} \;\; \sum_{k=1}^K R_{k}, \label{eq:P22a}\\
\text{s.t.}\quad
&R_{k\to j} \ge R_{k},\; \Omega(j)>\Omega(k), \label{eq:P22b}\\
&R_k \ge R_k^{\mathrm{min}}, \forall k\in \mathcal{K},\label{eq:P22c} \\
&\sum_{k=1}^{K} \|\mathbf{p}\| \le P_T. \label{eq:P22d}
\end{align}
\end{subequations}
This subproblem has a non-concave objective function and a non-convex constraint \eqref{eq:P22b}. To solve it, we define the following auxiliary variable \cite{beamforming}: 
\begin{equation}
    Q_k = \frac{1}{\sigma^2}\mathbf{p}\mathbf{p}^T\sum_{j = \Omega(k)}^{\Omega(K)}\alpha_j,
\end{equation}
As such, the objective can be rewritten as:
\begin{align}
    R_{\mathrm{sum}} = &\sum_{k=1}^{K-1}\log\left(\frac{\sum_{j = \Omega(k)}^{\Omega(K)}|\mathbf{h}_k\mathbf{p}^T|^2\alpha_j +\sigma^2}{\sum_{j = \Omega(k+1)}^{\Omega(K)}|\mathbf{h}_k\mathbf{p}^T|^2\alpha_j +\sigma^2}\right) \notag\\
    &\qquad\qquad\qquad\qquad\quad+\log(1 + \mathbf{h}_K^H\mathbf{Q_N}\mathbf{h}_K), \notag\\
    &= \sum_{k=1}^{K-1}\left(\log(1+\mathbf{h}_k^H\mathbf{Q_k}\mathbf{h}_k)-\log(\mathbf{h}_k^H\mathbf{Q_{k+1}}\mathbf{h}_k)\right) \notag\\
    &+ \log(1 + \mathbf{h}_K^H\mathbf{Q_N}\mathbf{h}_K) \notag= \sum_{k = 1}^{K}{\mathcal{F}}_k(\mathbf{Q}_k),
\end{align}
where
\begin{equation}
    \mathcal{F}_k(\mathbf{Q}_k) = \log(1+\mathbf{h}_k^H\mathbf{Q_k}\mathbf{h}_k)-\log(\mathbf{h}_{k-1}^H\mathbf{Q_{k}}\mathbf{h}_{k-1})
\end{equation}
for $k = 2,\dots K$ and $\mathcal{F}_1(\mathbf{Q}_1) = \log(1+\mathbf{h}_1^H\mathbf{Q_1}\mathbf{h}_1)$. Similarly, constraint \eqref{eq:P22b} is transformed into the equivalent form:
\begin{align}
    &C1:\mathcal{G}_{k,j}(\mathbf{Q}_k\mathbf{Q}_{k+1}) = \notag \\ &\log\biggl(\frac{\mathbf{h}_j^H\mathbf{Q}_k\mathbf{h}_j+1}{\mathbf{h}_j^H\mathbf{Q}_{k+1}\mathbf{h}_j+1}\biggl)-\log\biggl(\frac{\mathbf{h}_k^H\mathbf{Q}_k\mathbf{h}_k+1}{\mathbf{h}_k^H\mathbf{Q}_{k+1}\mathbf{h}_k+1}\biggl) \ge 0, \notag \\ & \qquad\qquad\qquad\qquad\qquad\qquad\qquad\qquad\qquad k,j \in \mathcal{M},
\end{align}
where $\mathcal{M}$ is the set of decoding constraint pairs. 
The QoS constraint \eqref{eq:P22c} can be linearized into:
\begin{align}
&C2: \mathbf{h}_k^H(\mathbf{Q}_k -a_k\mathbf{Q}_{k+1})\mathbf{h}_k + 1-a_k \ge 0, \notag \\ & \hspace{4cm}
\forall k = (1,2 \dots K-1), \\
&C3: \mathbf{h}_K^H\mathbf{Q}_K\mathbf{h}_K \ge a_N -1, 
\end{align}
where $a_k = 2^{R_k^{\mathrm{min}}}$. Consequently, problem (P2.2) can be equivalently transformed into the following problem:
\begin{align*}
\text{(P2.3):}\quad
&\max_{\boldsymbol{Q}} \;\; \sum_{k=1}^K \mathcal{F_k}(\mathbf{Q}_k), \\
\text{s.t.}\quad
&\eqref{eq:P22d}, C1, C2, C3,\\
&C4: \mathbf{Q}_k \succeq 0, k \in \mathcal{K}.\\
&C5: Q_k = \frac{1}{\sigma^2}\mathbf{p}\mathbf{p}^T\sum_{j = \Omega(k)}^{\Omega(K)}\alpha_j,
\end{align*}
Note that (P2.3) is still a non-convex problem, as the objective function is not concave and constraints C1 and C5 are not convex. So, following \cite{beamforming}, the lower bound of $\mathcal{F}_k(\mathbf{Q}_k)$ is introduced and, at any feasible $\mathbf{Q}^0_K$ for $k = 2, \dots, K$, it is:
\begin{align}
&\mathcal{F'}_k(\mathbf{Q}_k,\mathbf{Q}_k^{(0)}) = \notag \\ &  \hspace{1cm}\log(1+\mathbf{h}_k^H\mathbf{Q}_k\mathbf{h}_k)-\text{Tr}(\mathbf{A}(\mathbf{Q}_k^0)\mathbf{Q}_k)-B(\mathbf{Q}^0_k),
\end{align}
where
\begin{align*}
    &\mathbf{A}(\mathbf{Q}_k^0) = \frac{1}{\ln2}\mathbf{h}_{k-1}(1+\mathbf{h}_{k-1}^H\mathbf{Q}_k^0\mathbf{h}_{k-1})^{-1}\mathbf{h}_{k-1}^H,\\
    &\mathbf{B}(\mathbf{Q}_k^0) = \log(\mathbf{h}_{k-1}(1+\mathbf{h}_{k-1}^H\mathbf{Q}_k^0\mathbf{h}_{k-1}) - \text{Tr}(\mathbf{A}(\mathbf{Q}_k^0)\mathbf{Q}_k),
\end{align*}
Also, a lower bound of $ \mathcal{G}_{k,j}(Q_k, Q_{k+1})$  at any feasible $ (Q_k^0, Q_{k+1}^0)$ is given by:
\begin{align}
\mathcal{G'}_{k,j}&(\mathbf{Q}_k, \mathbf{Q}_{k+1}, \mathbf{Q}_k^{(0)}, \mathbf{Q}_{k+1}^{(0)}) 
= \log\big(h_j^H Q_k h_j + 1\big) \nonumber\\
&+\log\big(h_k^H Q_{k+1} h_k + 1\big) \nonumber \\
& - \mathrm{Tr}\big(E_{k,j}(Q_{k+1}^0) Q_{k+1} + F_{k,j}(Q_k^0) Q_k \big) \\
&- D_{k,j}(Q_k^0, Q_{k+1}^0), \quad \forall k,j \in \mathcal{K} \nonumber
\end{align}
where
\begin{align*}
&E_{k,j}(Q_{k+1}^0) = \frac{h_j h_j^H}{\ln 2 \,(1 + h_j^H Q_{k+1}^0 h_j)}, \\
&F_{k,j}(Q_k^0) = \frac{h_k h_k^H}{\ln 2 \,(1 + h_k^H Q_k^0 h_k)}, \\
&D_{k,j}(Q_k^0, Q_{k+1}^0) 
= \log(h_j^H Q_k^0 h_j + 1) + \log(h_k^H Q_{k+1}^0 h_k + 1) \nonumber \\
&\quad\>\> - \mathrm{Tr}\big(E_{k,j}(Q_{k+1}^0) Q_{k+1}^0 + F_{k,j}(Q_k^0) Q_k^0\big).
\end{align*}
Note that the approximated problem (P2.3) is still non-convex due to C5. Therefore, Successive Convex Approximation (SCA) is adopted to obtain a near-optimal solution. At iteration $t$, given a feasible point $\mathbf{Q}_k^{(t)}$, SCA replaces the non-concave terms in both the objective and the constraints with their concave lower bounds derived above. As for the rank-$1$ constraint C5, it is linearized around a feasible point $\mathbf{p}^{(t)}$ at iteration $t$. Using first-order Taylor expansion:
\begin{equation}
    \mathbf{p}\mathbf{p}^T \approx \mathbf{p}^{(t)}(\mathbf{p}^{(t)})^T + \mathbf{p}^{(t)}(\mathbf{p} - \mathbf{p}^{(t)})^T + (\mathbf{p} - \mathbf{p}^{(t)})(\mathbf{p}^{(t)})^T.
\end{equation}
Simplifying (21) leads to: 
\begin{equation}
    \mathbf{p}\mathbf{p}^T \approx 2\mathbf{p}^{(t)}(\mathbf{p})^T - \mathbf{p}^{(t)}(\mathbf{p}^{(t)})^T.
\end{equation}
Therefore, constraint C5 can be approximated with:
\begin{equation}
    \mathbf{Q}_k \approx \frac{1}{\sigma^2}\left(2\mathbf{p}^{(t)}(\mathbf{p})^T - \mathbf{p}^{(t)}(\mathbf{p}^{(t)})^T\right)\sum_{j = \Omega(k)}^{\Omega(K)}\alpha_j,
\end{equation}
and, can be rewritten as:
\begin{equation*}
    C5': \mathbf{Q}_k = \frac{2}{\sigma^2}\mathbf{p}^{(t)}(\mathbf{p})^T\sum_{j = \Omega(k)}^{\Omega(K)}\alpha_j - \frac{1}{\sigma^2}\mathbf{p}^{(t)}(\mathbf{p}^{(t)})^T\sum_{j = \Omega(k)}^{\Omega(K)}\alpha_j.
\end{equation*}
\textit{SCA Iterative Algorithm}: At iteration $t$, given a feasible point $(\mathbf{Q}_k^{(t)}, \mathbf{p}^{(t)})$ for $k \in \mathcal{K}$, Algorithm 1 solves the following convex approximation problem:
\begin{subequations}\label{eq:P2.4}
\begin{align*}
\text{(P2.4):}\quad
&\max_{\boldsymbol{Q}, \mathbf{p}} \;\; \sum_{k=2}^K \mathcal{F}'_k(\mathbf{Q}_k, \mathbf{Q}_k^{(t)}) + \mathcal{F}_1(\mathbf{Q}_1) \label{eq:P24a}\\
\text{s.t.}\quad
&C1':\mathcal{G'}_{k,j}(\mathbf{Q}_k, \mathbf{Q}_{k+1}, \mathbf{Q}_k^{(t)}, \mathbf{Q}_{k+1}^{(t)}) \ge 0, \nonumber\\
&\quad\quad\quad\quad\quad\quad\quad k,j \in \mathcal{K}, \Omega(j)>\Omega(k) \\
&\eqref{eq:P22d}, C2, C3, C5'.
\end{align*}
\end{subequations}
Problem P2.4 is a convex optimization problem that can be efficiently solved using standard convex optimization solvers such as CVX. The algorithm is summarized in Algorithm 1.

\begin{algorithm}[!t]
\caption{Alternating Optimization for Joint Power Allocation}
\begin{algorithmic}[1]
\STATE \textbf{Initialize:} Set $\mathbf{p}^{(0)}$, $\boldsymbol{\alpha}^{(0)}$, iteration index $\tau = 0$, tolerance $\epsilon > 0$
\STATE Compute initial sum rate $R_{sum}^{(0)}$
\REPEAT
    \STATE \textbf{Step 1: User Power Allocation}
    \STATE Fix $\mathbf{p} = \mathbf{p}^{(\tau)}$
    \STATE Determine SIC ordering $\Omega$ based on $|\mathbf{h}_k\mathbf{p}^{(\tau)}|^2$
    \STATE Solve for $\boldsymbol{\alpha}^{(\tau+1)}$ using closed-form expressions in (14)-(15)
    \STATE \textbf{Step 2: Radiation Power Allocation via SCA}
    \STATE Fix $\boldsymbol{\alpha} = \boldsymbol{\alpha}^{(\tau+1)}$
    \STATE Initialize SCA: $t = 0$, $\mathbf{p}^{SCA,(0)} = \mathbf{p}^{(\tau)}$
    \REPEAT
        \STATE Compute $\mathbf{Q}_k^{(t)}$ from $\mathbf{p}^{SCA,(t)}$ and $\boldsymbol{\alpha}^{(\tau+1)}$
        \STATE Solve convex problem P2.4 to obtain $\mathbf{p}^{SCA,(t+1)}$
        \STATE $t \leftarrow t + 1$
    \UNTIL{Convergence}
    \STATE $\mathbf{p}^{(\tau+1)} = \mathbf{p}^{SCA,(t)}$
    \STATE Compute sum rate $R_{sum}^{(\tau+1)}$
    \STATE $\tau \leftarrow \tau + 1$
\UNTIL{$|R_{sum}^{(\tau)} - R_{sum}^{(\tau-1)}| < \epsilon$}
\STATE \textbf{Output:} Optimal power allocation $(\mathbf{p}^*, \boldsymbol{\alpha}^*)$
\end{algorithmic}
\end{algorithm}

\begin{figure*}[!t]
 \centering
  \begin{subfigure}[t]{0.32\textwidth}
    \includegraphics[width=\linewidth]{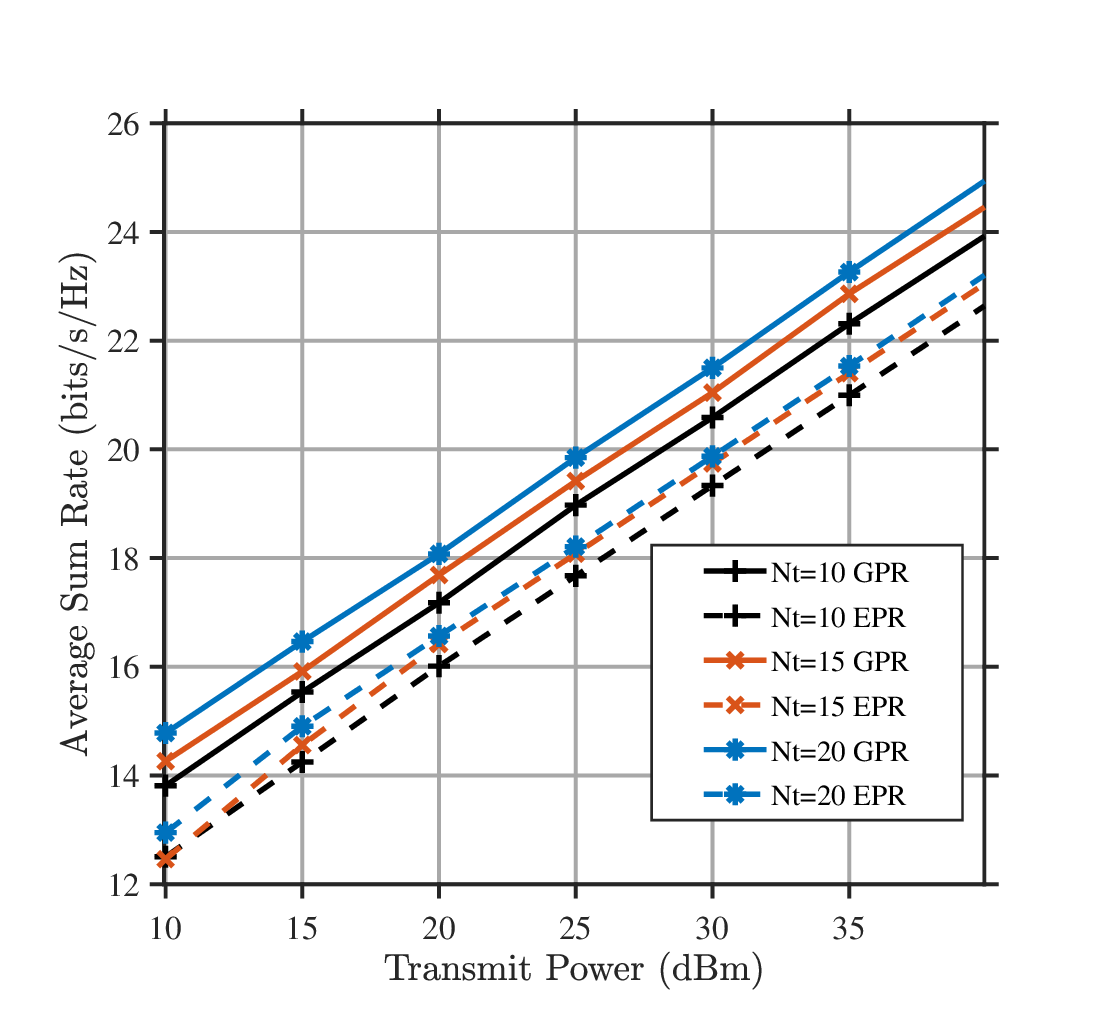}
    \caption{}\vspace{-0.3cm}
    \label{fig:Vary_antenna}
  \end{subfigure}
  \begin{subfigure}[t]{0.32\textwidth}
     \includegraphics[width=\linewidth]{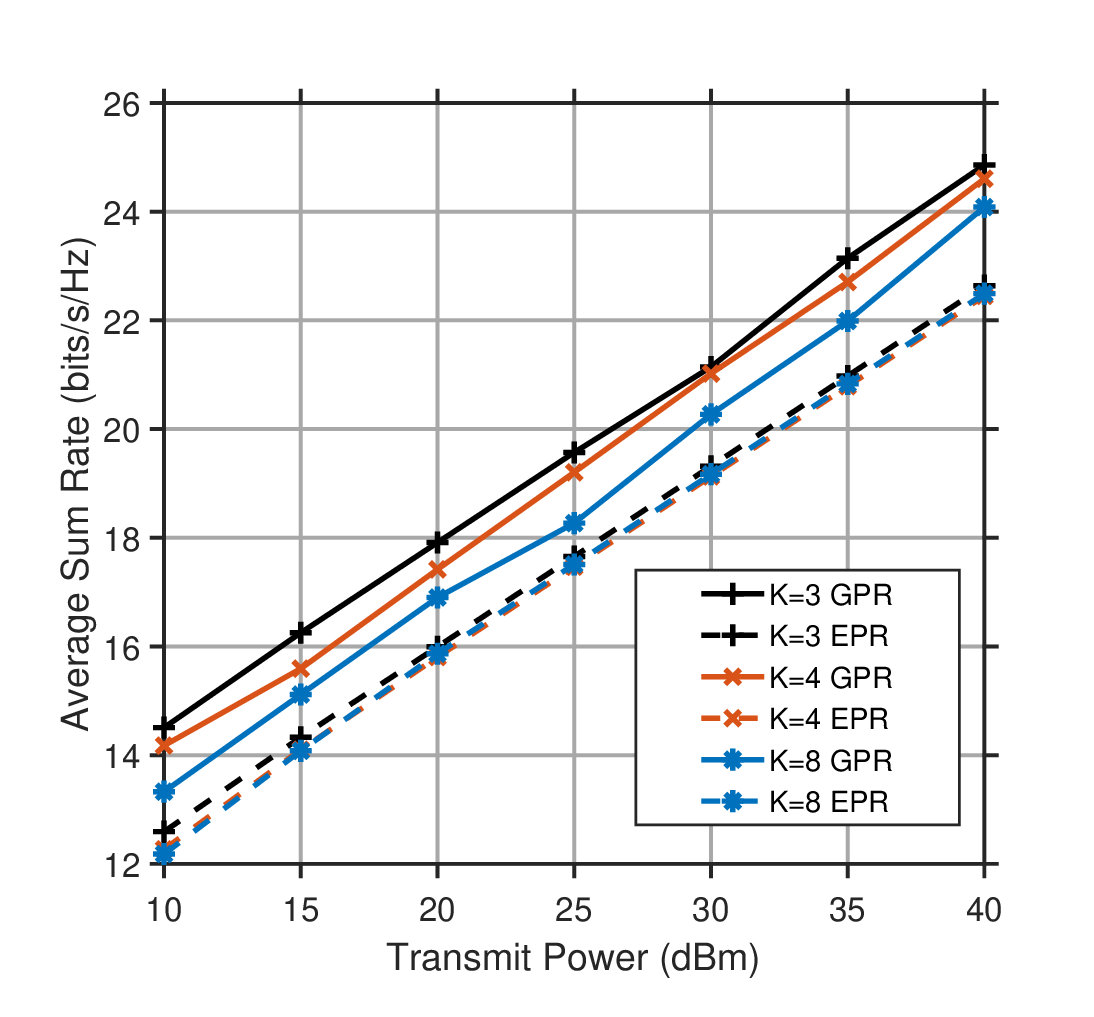}
     \caption{} \vspace{-0.3cm}
     \label{fig:Vary_user}
  \end{subfigure}
  \begin{subfigure}[t]{0.32\textwidth}
     \includegraphics[width=\linewidth]{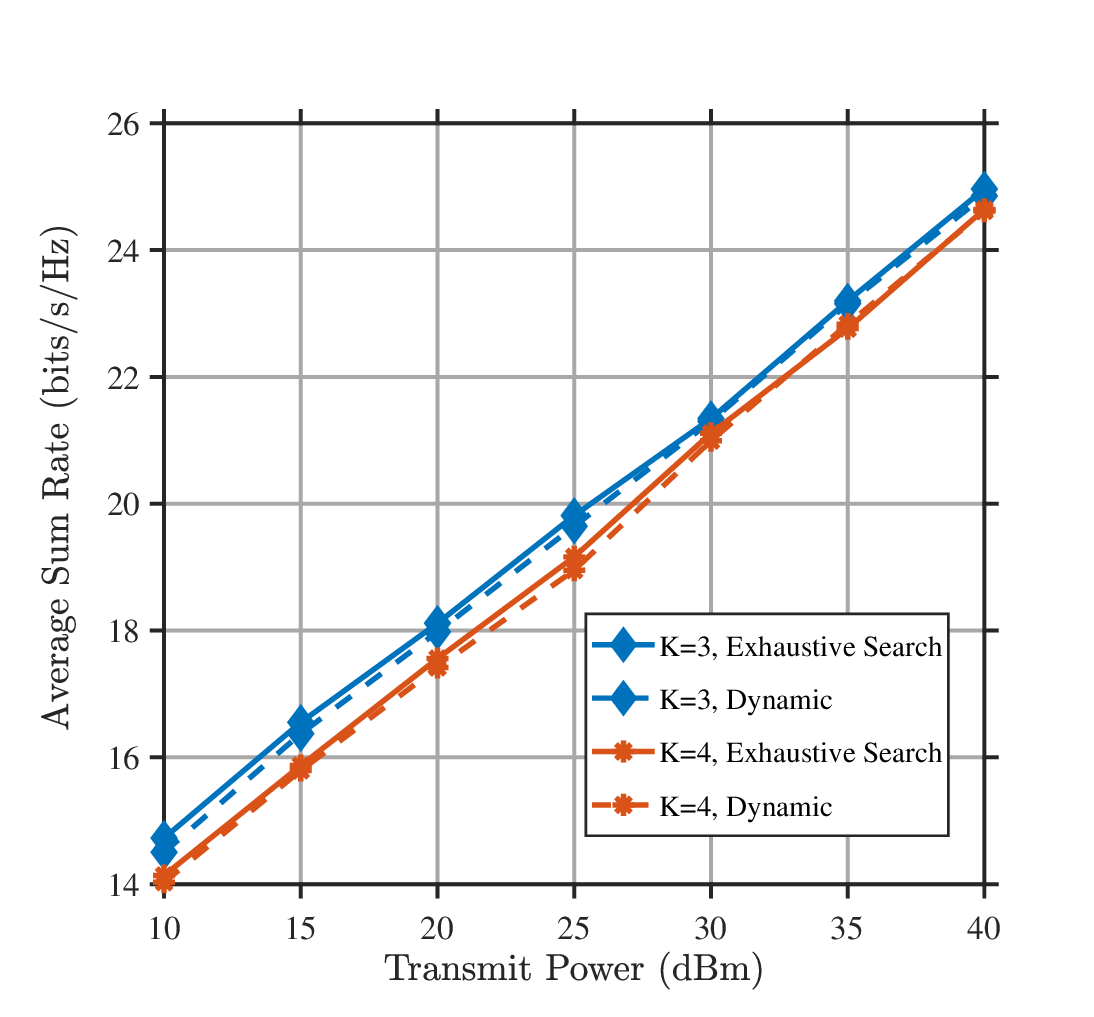}
     \caption{}\vspace{-0.3cm}
     \label{fig:FigC}
  \end{subfigure}
  \caption{(a) Sum rate of a general power model system and equal power model system for $K=3$ users; (b) Sum rate comparison between general power model and equal power model systems for varying number of users and $N_t = 20$; (c) Comparison of sum rates for dynamically determined and exhaustive search-based SIC ordering.}
  \vspace{-0.4cm}
\end{figure*}
\vspace{-0.3cm}
\section{Numerical Analysis and Simulations}
To assess the effectiveness of the proposed joint power allocation and radiation optimization framework, extensive simulations were conducted using a custom-built MATLAB-based environment implementing the mathematical formulations derived in Section IV. The considered PASS-assisted downlink NOMA system operates at a carrier frequency of $28$ GHz, with the BS positioned at a height of $3$ m and serving $K$ single-antenna users uniformly distributed within a $6 \times 10$ $\text{m}^2$ rectangular area. The dielectric waveguide is equipped with $N_t$ equally spaced PAs located along its length, and the effective refractive index of the waveguide is set to $n_\text{eff} = 1.4$. The AWGN variance is fixed at $\sigma^2 = -90$ dBm. The minimum QoS requirement for all users is $R_k^\text{min} = 0.5$ bit/s/Hz. The BS' total transmitted power is normalized to unity. For benchmarking, a conventional NOMA-assisted PASS architecture employing the EPR model and antenna activation strategy in \cite{AntennaActivation} is simulated under identical conditions. The proposed scheme herein, jointly optimizes both user-side NOMA power allocation and PA radiation coefficients via the earlier-described AO and SCA algorithms. Each result is averaged over $100$ independent channel realizations to ensure statistical reliability.

Fig.\ref{fig:Vary_antenna} illustrates the variation of the system's achievable sum rate under both the GPR model and the benchmark EPR configuration, for different numbers of PAs ($N_t$). The results reveal a consistent improvement in sum rate as $N_t$ increases, since a denser PA deployment enhances spatial diversity and strengthens the effective LoS channels between the BS and users. However, while both systems benefit from additional antennas, the gain realized by the proposed GPR-based scheme is notably larger. This improvement stems from the additional degree of freedom introduced by the controllable radiation coefficients, which allows the optimizer to tailor the emitted power from each PA according to instantaneous channel conditions and decoding requirements. As $N_t$ grows, the EPR configuration begins to exhibit diminishing returns, primarily because equal power distribution cannot exploit the spatial selectivity of the individual PAs once the waveguide becomes densely populated. In contrast, the GPR scheme maintains a steady sum-rate gain even at higher PA densities, as the optimization framework effectively utilizes the richer antenna granularity to refine the overall radiation pattern and mitigate inter-user interference. These results clearly demonstrate the scalability and robustness of the proposed approach in multi-antenna PASS deployments. 


Fig.\ref{fig:Vary_user} examines the achievable sum rate performance of both the proposed GPR-based system and the EPR benchmark under different numbers of downlink users ($K$) with $N_t = 20$ PAs. The total sum rate gradually decreases as the number of users increases; a trend primarily attributed to the minimum QoS requirement imposed by each user ($R_k^\text{min} = 0.5$ bit/s/Hz). In multi-user NOMA systems, a larger $K$ inherently introduces additional decoding constraints; hence, forcing the optimizer to allocate a greater fraction of the total transmit power to weaker users to ensure fairness and guarantee their QoS targets (this is inline with \cite{OptimalSIC}). Consequently, less power remains available for stronger users (\textit{i.e.}, those with more favorable channel conditions); thereby, reducing the overall system throughput. Nevertheless, across all tested user configurations, the proposed GPR-based approach consistently outperforms the EPR-based counterpart. This performance superiority arises from the ability of the GPR model to dynamically shape the radiation pattern of each PA, which enhances the effective channel gain of the dominant users without violating QoS constraints. The flexible power coupling between PAs and users, thus, enables more efficient exploitation of NOMA's power domain multiplexing; hence, resulting in a significantly higher aggregate data rate even as network density increases. 


 Fig.\ref{fig:FigC} compares the sum-rate performance obtained using two different approaches for determining the SIC ordering, namely: \textit{i}) the high-complexity exhaustive search, and, \textit{ii}) the dynamic SIC update. In the dynamic approach, before solving subproblem (P2.1), the algorithm initializes the radiation power vector $\mathbf{p}$ and computes the corresponding effective channels $\textbf{h}_j\mathbf{p}^T$. The initial SIC order is then established according to the users' effective channel strengths, as discussed in Section IV-A. After each AO iteration, once the new $\mathbf{p}$ values are obtained, the SIC order is recalculated to reflect the updated channel conditions. Conversely, in the exhaustive search-based approach, the SIC ordering is determined once at the beginning and remains fixed throughout the optimization process. The results in Fig.\ref{fig:FigC} show that both schemes achieve nearly identical sum-rate performance; hence, indicating that the dynamic update mechanism effectively captures the optimal decoding sequence without the need for exhaustive search. Given its substantially lower computational cost, the dynamic SIC-ordering strategy provides an appealing balance between algorithmic efficiency and near-optimal performance. This is why it is particularly suitable for practical real-time PASS-NOMA implementations.


\section{Conclusion}

This paper studied a downlink NOMA-assisted PASS that maximized the sum rate by jointly optimizing user power allocation and PA radiation coefficients. The resulting non-convex optimization problem was decomposed into two subproblems and efficiently solved using AO, combining a closed-form solution for user power allocation with an SCA-based update for PA radiation powers. Simulation results confirmed the superiority of the proposed general power radiation model over equal-power schemes and showed that dynamically updating the SIC ordering achieves near-optimal performance with significantly lower complexity.

\vspace{12pt}
\bibliographystyle{IEEEtran}
\bibliography{IEEEabrv,reference}

@ARTICLE{10904090,
  author={Liu, Ruiqi and Zhang, Leyi and Li, Ruyue Yu-Ngok and Renzo, Marco Di},
  journal={IEEE Vehicular Technology Magazine}, 
  title={{The ITU Vision and Framework for 6G: Scenarios, Capabilities, and Enablers}}, 
  year={2025},
  volume={20},
  number={2},
  pages={114-122},
  doi={10.1109/MVT.2025.3532887}}

@ARTICLE{10054381,
  author={Wang, Cheng-Xiang and You, Xiaohu and Gao, Xiqi and Zhu, Xiuming and Li, Zixin and Zhang, Chuan and Wang, Haiming and Huang, Yongming and Chen, Yunfei and Haas, Harald and Thompson, John S. and Larsson, Erik G. and Renzo, Marco Di and Tong, Wen and Zhu, Peiying and Shen, Xuemin and Poor, H. Vincent and Hanzo, Lajos},
  journal={IEEE Communications Surveys \& Tutorials}, 
  title={{On the Road to 6G: Visions, Requirements, Key Technologies, and Testbeds}}, 
  year={2023},
  volume={25},
  number={2},
  pages={905-974},
  doi={10.1109/COMST.2023.3249835}}

@ARTICLE{11169486,
  author={Liu, Yuanwei and Wang, Zhaolin and Mu, Xidong and Ouyang, Chongjun and Xu, Xiaoxia and Ding, Zhiguo},
  journal={IEEE Communications Magazine}, 
  title={{Pinching-Antenna Systems: Architecture Designs, Opportunities, and Outlook}}, 
  year={2025},
  volume={},
  number={},
  pages={1-7},
  doi={10.1109/MCOM.001.2500037}}

@article{liu2025pinchingTutorial,
  title={{Pinching-Antenna SyStems (PASS): A Tutorial}},
  author={Liu, Yuanwei and Jiang, Hao and Xu, Xiaoxia and Wang, Zhaolin and Guo, Jia and Ouyang, Chongjun and Mu, Xidong and Ding, Zhiguo and Nallanathan, Arumugam and Karagiannidis, George K and others},
  journal={arXiv preprint arXiv:2508.07572},
  year={2025}
}

@ARTICLE{11215676,
  author={Zeng, Ming and Wang, Ji and Li, Xingwang and Wang, Gongpu and Dobre, Octavia A. and Ding, Zhiguo},
  journal={IEEE Wireless Communications Letters}, 
  title={{Sum Rate Maximization for NOMA-Assisted Uplink Pinching-Antenna Systems}}, 
  year={2025},
  volume={},
  number={},
  pages={1-1},
  doi={10.1109/LWC.2025.3624761}}

@ARTICLE{AntennaActivation,
  author={Wang, Kaidi and Ding, Zhiguo and Schober, Robert},
  journal={IEEE Wireless Communications Letters}, 
  title={{Antenna Activation for NOMA Assisted Pinching-Antenna Systems}}, 
  year={2025},
  volume={14},
  number={5},
  pages={1526-1530},
  doi={10.1109/LWC.2025.3548280}}

@ARTICLE{OptimalSIC,
  author={Rezvani, Sepehr and Jorswieck, Eduard Axel and Yamchi, Nader Mokari and Javan, Mohammad Reza},
  journal={IEEE Transactions on Wireless Communications}, 
  title={{Optimal SIC Ordering and Power Allocation in Downlink Multi-Cell NOMA Systems}}, 
  year={2022},
  volume={21},
  number={6},
  pages={3553-3569},
  doi={10.1109/TWC.2021.3120325}}

@ARTICLE{beamforming,
  author={Zhu, Fusheng and Lu, Zhaohua and Zhu, Jianyue and Wang, Jiaheng and Huang, Yongming},
  journal={IEEE Access}, 
  title={Beamforming Design for Downlink Non-Orthogonal Multiple Access Systems}, 
  year={2018},
  volume={6},
  number={},
  pages={10956-10965},
  doi={10.1109/ACCESS.2018.2797209}}

\end{document}